\documentstyle[12pt]{article}
\begin{document}
\centerline{\bf Space covering by growing rays}
\bigskip\bigskip
\centerline{\bf P.~L.~Krapivsky$^1$ and E.~Ben-Naim$^2$}
\bigskip
\centerline{\sl $^1$Courant Institute of Mathematical Sciences,
New York University, New York, NY 10012}
\centerline{\sl $^2$The James Franck Institute, The University of Chicago,
Chicago, IL 60637}

\bigskip\bigskip

\begin{abstract}\noindent
We study kinetic and jamming properties of a space covering process in
one dimension. The stochastic process is defined as follows: Seeds are
nucleated randomly in space and produce rays which grow with a
constant velocity.  The growth stops upon collision with another
ray. For arbitrary distributions of the growth velocity, the exact
coverage, velocity and size distributions are evaluated for both
simultaneous and continuous nucleation. In general, simultaneous
nucleation exhibits a stronger dependence on the details of the growth
velocity distribution in the asymptotic time regime. The coverage in
the continuous case exhibits a universal $t^{-1}$ approach to the
jammed state, while an inhomogeneous version of the process leads
to nonuniversal $t^{-p_+}$ decay, with $0\le p_+\le 1$ the fraction of
right growing rays.
\end{abstract}
\vskip .6in

\section{Introduction}

Covering of space by growing objects occurs in numerous natural
phenomena; phase separation \cite{Bray}, phase transformation
\cite{Sekimoto}, monolayer \cite{Evans, Meakin} and multilayer
\cite{Privman} adsorption, aggregation \cite{Viscek}, and wetting
\cite{sa} are just few examples.  Covering processes can differ by the
input mechanism, the covering object growth dynamics, and the
interaction between the objects.  In reality, collisions may proceed
through various physico-chemical processes so the description of a
single collision event can be a difficult task.  However, in the realm
of statistical mechanics one seeks to describe the essential mechanism
that underlies the collective behavior, thereby illuminating the
complexity resulting from the many-body nature of the problem.
Therefore, in this study we will consider models with the simplest
possible input rules (all seeds nucleate simultaneously or
continuously with constant rate), the simplest ballistic growth law,
and a simple collision rule - when an object hits another object, it
stops.

This simple model mimics various physical, chemical, and biological
processes. For example, a reacted functional group in a 1D polymer
chain can poison adjacent unreacted group, this new reacted group then
poison next, etc. In many situations the reaction may be {\it
directed} thus giving an example of the ray model we will examine
below. Another natural example is a directed biological growth and
spreading \cite{Jager}.

Several recent theoretical works model the covering process using
ballistically growing objects, see e.g. \cite{Sekimoto, Evans, Meakin,
Bradley, Andrienko, Brilliantov} and references therein.  In other
studies (see e.g. \cite{Krapivsky, india}), the dimension of the objects
is smaller than the dimension of the space and hence in such processes
space splitting  rather than space covering occurs.  We study
the growth of rays from point seeds, a stochastic process which
results in space covering in one dimension.  The process proceeds as
follows: Rays grow freely with velocity $v$ until they hit other rays;
after a collision of a moving tip with another ray, the tip stops and
that ray becomes frozen. There are two natural nucleation rules:
homogeneous and hetereogeneous. In the heterogeneous model, seeds are
nucleated simultaneously. We consider the simplest case of initially
uncorrelated spatial distribution of seeds.  The velocity is
distributed according to an arbitrary velocity distribution $P_0(v)$.  For the
homogeneous model, the nucleation process also starts at $t=0$ but
proceeds forever: Seeds are nucleated stochastically {\it both} in
space and in time, with a constant rate per unit volume of uncovered space.
These two cases have a significant difference -- while the homogeneous
process is stochastic, the heterogeneous process is deterministic.
The spatial dimension $d$ plays an important role in these growth
processes.  For $d=1$, growing rays cover a finite fraction of the
space, eventually all the space for the homogeneous model.  The
homogeneous model has some similarity to the Avrami-Kolmogorov
nucleation-and-growth process \cite{Sekimoto}, while the heterogeneous
one resembles some properties of random sequential adsorption
\cite{Evans}.  In contrast, for $d>1$ the net volume covered by rays
is zero, hence some characteristics can be computed trivially,
e.g. the number density of rays is equal to $t$, while the geometric
patterns are rich and interesting.

In the following we obtain exact results for arbitrary velocity
distribution of the input in one-dimension. Generally, a jamming
configuration is approached for long times and we are interested in
both the kinetic and the jamming properties.  We focus on simple
quantities such as the coverage, the velocity distribution of growing
rays and the size distribution of the rays in the jammed
configuration.  The heterogeneous case is considered in section II,
and the homogeneous case is considered in section III. We briefly
discuss a possible generalization to higher dimensions and describe it
using a mean-field technique in section IV.

\section{Heterogeneous Nucleation}

In the heterogeneous case seeds are randomly distributed on a line with
concentration $c_0$ at time $t=0$. Each seed gives birth to a ray
whose tip moves freely with a constant intrinsic velocity. When the
tip of a ray collides with either another seed or another tip, its
growth is stopped. The growth velocities are independent of the
position and are distributed according to $P_0(v)$, such that
$c_0=\int dv P_0(v)$.  We assume that the velocity distribution has a
characteristic velocity $v_0$, {\it i.e.}, it can be written as
$(c_0/v_0)P_0(v/v_0)$ with $\int dz P_0(z)=1$.  It is conveneient make
a transformation to dimensionless time $c_0v_0t\to t$, velocity
$v/v_0\to v$, concentration $c/c_0\to c$, and velocity distribution
$(c_0/v_0)P_0\to P_0$.  Thus, the initial seed concentration is set to
unity.  In the following we obtain the exact jamming properties of the
system as well as its approach towards the jammed state for arbitrary
velocity distribution.  We then consider the behavior for two special
cases, a bimodal velocity distribution and a distribution with a
power-law behavior in the small-velocity limit.

Several properties of the jammed configuration can be easily derived
from the initial state.  Let us introduce the shorthand notations for
the velocity distribution and the density of right- and left-moving
rays, $P_{\pm}(v)=P_0(v)\theta(\pm v)$, with $\theta(x)$ the Heaviside
step function, and $p_{\pm}=\int dv P_{\pm}(v)$, which in turn implies
$p_++p_-=1$.  The final fraction of uncovered space is
\begin{equation}
\Phi_{\infty}=p_+p_-,
\end{equation}
This follows from a simple observation: For a pair of adjacent seeds,
the space between them remains completely uncovered if the left ray
moves to the left and the right one moves to the right; for the three
other situations, the space between the adjacent seeds will be
covered. Of course, the final coverage is given by $1-\Phi$, with
$3/4\leq 1-\Phi\leq 1$. Maximal final coverage is achieved when all
rays grow in the same directions, while minimal final coverage occurs
for $p_+=p_-=1/2$.

One can solve for the kinetics of the covering process by a number of
techniques.  We will use an approach that proves powerful for the more
difficult problem of homogeneous nucleation.  This procedure has been
applied to few other nucleation-and-growth processes \cite{Bradley,
Andrienko, Brilliantov}.  We start by noting that the fraction
$\Phi(t)$ of uncovered space can be thought as the probability that
some point, say the origin, remains uncovered at time $t$.  The key
point of the approach is very simple: One first investigates an
auxiliary ``one-sided'' problem in which seeds are scattered to the
left of the origin and no seeds are placed to the right. Having
computed the probability $\phi_+(t)$ that the origin remains uncovered
up to time $t$ in this one-sided problem, the ``two-sided''
probability $\Phi(t)$ follows from a clear identity:
\begin{equation}
\Phi(t)=\phi_+(t)\phi_-(t),
\end{equation}
where $\phi_-(t)$ corresponds to the complementary one-sided problem.
To determine $\phi_+(t)$ we note that the origin is covered during the
time interval $(t,t+dt)$ with probability $(-d\phi_+/dt)dt$. On the
other hand, it happens if the nearest to the origin seed has a
positive velocity, say $v$, and lies in the space interval
$(-vt-vdt,-vt)$. Integrating over all positive velocities, we find
$(d\phi_+/dt)dt=-\int dv vdt P_+(v) e^{-vt}$. This leads to the rate
equation
\begin{equation}
{d\phi_+\over dt}=-\int dv v P_+(v) e^{-vt},
\end{equation}
with the initial condition $\phi_+(0)=1$. Solving Eq.~(3)
subject to this initial condition gives
\begin{equation}
\phi_+(t)=p_- + \int dv P_+(v) e^{-vt}.
\end{equation}
Similar expression holds for $\phi_-$. Combining
these two one-sided problems, we find
\begin{equation}
\Phi(t)=
\left(p_-+\int dv P_+(v)e^{-vt}\right)
\left(p_++\int dv P_-(v)e^{vt}\right).
\end{equation}
Indeed, the final uncovered fraction of space agrees with Eq.~(1).

For sufficiently small times, the uncovered fraction decreases linearly
with time according to
\begin{equation}
\Phi(t)\cong 1-\langle |v|\rangle_0 t, \qquad t\to 0.
\end{equation}
The prefactor is equal to the average ray speed, and in the following
$\langle\cdots\rangle_0\equiv\int dv\langle\cdots\rangle P_0(v)$.
Initially, the rays cover the space very effectively, and as the
process continues, the overall covering rate decreases depending on
the nature of the initial velocity distribution.  In the long time
limit, the fraction of uncovered space approaches its final value
according to
\begin{equation}
\Phi(t)-\Phi_{\infty}\sim \int_0^\infty dv
\tilde P(v) e^{-vt}\qquad t\to\infty,
\end{equation}
with the modified velocity distribution
$\tilde P(v)=p_+P_+(v)+p_-P_-(-v)$.  The above integral is simply the Laplace
transform of $\tilde P(v)$, and in the long time limit it is
dominated by the velocity distribution near the minimal velocity of
the distribution.  We conclude that slow rays dominate asymptotically,
as demonstrated below for a distribution which
behaves as a power-law near the origin.

To determine the size distribution, we first compute $P(v,t)$, the
density of  rays of velocity $v$ that have not been stopped before
time $t$. This density is simply given by
\begin{equation}
P_{\pm}(v,t)=P_{\pm}(v,0)e^{-|v|t}\phi_{\mp}(t).
\end{equation}
Here the exponential factor gives the probability that the interval of
length $|v|t$, covered by the growing ray at time $t$, does not contain
other seeds; the latter factor ensures that the point reached by the
growing tip at time $t$, remains uncovered by rays growing from the
other half-space.  The covering rate, $-d\Phi/dt$, can be calculated
from the velocity distribution, $-d\Phi(t)/dt=\int dv |v| P(v,t)$.
This result is consistent with the exact solution of Eq.~(5).

Eq.~(8) allows us to find the density of frozen rays of length $l$ in
the final (jammed) configuration, $\rho_\infty(l)$.  Frozen rays of
length $l$ are those that stopped their growth at time
$t=l/|v|$. Therefore, the limiting size distribution is related to the
velocity distribution via $\rho_\infty(l)=-\int dt\int dv (\partial
P(v,t)/\partial t)\delta(l-|v|t)$.  Substituting Eq.~(8) and
evaluating the integrals gives
\begin{eqnarray}
\rho_{\infty}(l)&=&(p_+^2+p_-^2)e^{-l}\nonumber\\
&+&\int dv\int du P_+(v)P_-(u)
\left[(1-u/v)e^{-l(1-u/v)}+(1-v/u)e^{-l(1-v/u)}\right].
\end{eqnarray}
The first term can be easily understood: Two adjacent rays moving in
similar directions and initially separated by distance $l$, give rise
to a frozen ray of length $l$. The exponential factor describes the
probability that such an interval is empty and the prefactor accounts
for the fraction of parallel moving neighbors. The second term
accounts for collisions between rays that grew in the opposite
directions.  One can verify by direct integration of Eq.~(9) that the
final fraction of covered space $\int dl l \rho_{\infty}(l)$ equals
the previously established value $1-p_+p_-$, thus providing a useful
check of consistency.

In the jammed state, clusters of frozen rays are separated by gaps.
Following the above calculations, it might be possible to find the
density of $n$-rays clusters of length $l$, and even more detailed
quantities. One such quantity is $\tilde\rho_{\infty}(l)$, the
distribution of gaps of size $l$ at the final state. This distribution
is easily found, $\tilde\rho_{\infty}(l)=p_-p_+e^{-l}$ and indeed, it
satisfies the normalization condition $\Phi_{\infty}=\int dl
l\tilde\rho_{\infty}(l)=p_+p_-$.

We discuss now  several specific initial velocity distributions.
We first consider the case of bimodal velocity distributions
\begin{equation}
P_0(v)=p_+\delta(v-1)+p_-\delta(v+1),
\end{equation}
with $p_++p_-=1$. The fraction of uncovered space is found from Eq.~(5),
\begin{equation}
\Phi(t)=(p_-+p_+e^{-t})(p_++p_-e^{-t})
\end{equation}
The approach towards the jammed state is a fast exponential one,
$\Phi(t)-\Phi_{\infty}\simeq (p_+^2+p_-^2)e^{-t}$ as $t\to\infty$.
Hence, the density of active rays is exponentially decreasing with
time as well. The jamming length distribution can be evaluated using
Eq.~(9),
\begin{equation}
\rho_{\infty}(l)=(p_+^2+p_-^2)e^{-l}+4p_+p_-e^{-2l}.
\end{equation}
The exponential asymptotic behavior holds for polydisperse velocity
distributions as long as the distribution vanishes in the vicinity of
the origin. To examine the effects of slow rays on the asymptotic
behavior, it is useful to consider power-law distributions
\begin{equation}
P_0(v)\sim |v|^{\mu}, \qquad {\rm when} \quad v\to 0,
\end{equation}
with $\mu>-1$. The long time
asymptotics is governed by the large argument Laplace transform of the
velocity distribution  and consequently  an algebraic
asymptotic behavior is found for the uncovered fraction:
\begin{equation}
\Phi(t)-\Phi_{\infty}\sim t^{-1-\mu},\qquad {\rm as} \quad t\to\infty.
\end{equation}
Furthermore, Eq.~(8) indicates that $P(v,t)\sim |v|^{\mu}\exp(-|v|t)$.
In other words, the velocity distribution can be written in a
scaling form
\begin{equation}
P(v,t)\sim t^{-\mu}f(z), \qquad {\rm with}\quad z=|v|/\langle |v|\rangle.
\end{equation}
In the above equation the typical velocity decays in time according to
$\langle |v|\rangle\sim t^{-1}$ and the scaling function is
$f(z)=z^{\mu}e^{-z}$.  The total density of growing rays at time $t$,
$n_r(t)$, decays according to $n_r\sim t^{-\mu-1}$.  To see how the
exponential behavior turns into a power-law one, we note that
$P_0(v)\sim \left(|v|-v_{\rm min}\right)^{\mu}$ for $|v|\to v_{\rm
min}$, with $v_{\rm min}>0$, leads to $\Phi(t)-\Phi_{\infty} \sim
t^{-\mu-1}e^{-v_{\rm min}t}$.  This situation, where the smallest
velocities dominate, is reminiscent of ballistic aggregation and
annihilation processes with continuous velocity distributions
\cite{Ben-Naim}.

\section{Homogeneous Nucleation}

In the homogeneous case seeds are nucleated stochastically {\it both}
in space and in time. At time $t=0$, the system is assumed to be
empty, seeds appear constantly with rate of $\gamma_0$ on uncovered
space, and eventually the system reaches complete coverage.  It is
again convenient to introduce dimensionless velocity $v/v_0\to v$,
space $x\sqrt{\gamma_0/v_0}\to x$, time $t\sqrt{\gamma_0 v_0}\to t$,
and input distribution function $(\gamma_0/v_0)P_0\to P_0$.  In the
following, we write the equations describing the coverage and the gap
distribution. Although we do not obtain a general explicit solution
for the coverage, an asymptotic analysis is carried. For arbitrary
distributions, the asymptotic coverage is independent of most details
of the input velocity distribution. We obtain explicit results for the
kinetic and the jamming properties in the case of bimodal velocity
distributions.

We again consider first the one-sided problem: seeds are deposited
only to the left of the origin.  Repeating the steps used in deriving
Eq.~(3), $\phi_+(t)$, the probability that the origin remains
uncovered at time $t$ satisfies
\begin{equation}
{d\phi_+\over dt}=-\int dv v P_+(v)\,e^{-vt^2/2}
\int_0^t d\tau\,e^{v\tau^2/2}\phi_+(\tau).
\end{equation}
Indeed, the origin can be covered during the time interval $(t,t+dt)$
by a $v$-velocity seed with positive direction of growth that could
have been nucleated at time $\tau$, with $0<\tau<t$, in the spatial
interval $\left(-v(t-\tau)-vdt,-v(t-\tau)\right)$.  Hence,
the integration over the velocity and the time variables.  The
exponential factor ensures that no nucleation have occurred in the
spatial interval covered by the ray before the seed appeared and that
no nucleation happens in the shrinking part of this interval during the
growth of the ray. The last factor ensures that the point $-v\tau$ is
uncovered at time $\tau$.  We have used  the one-sided probability
$\phi_+(\tau)$ since the condition that at time $\tau$ the point
$-v\tau$ was not covered from the right has already been taken into
account by the exponential factor.  Despite the complex structure of
this rate equation, the most interesting aspects of the covering
process, {\it i.e.} the short and long time behavior, can be found for
an arbitrary distribution without an explicit solution.

The early behavior of the system can be easily found by setting the
second integral to $t$, and consequently $\phi_{\pm}(t)\cong
1-B_{\pm}t^2/2$ with $B_{\pm}=\int dv |v| P_{\pm}(v)$. Hence, the early
time behavior is given by
\begin{equation}
\Phi(t)\cong 1-{\langle|v|\rangle_0\over 2} t^2\qquad t\to 0.
\end{equation}
This initial coverage is slower than in the homogeneous case since no
rays are initially present.  Note that in both cases the prefactor
equals the average ray speed.

We turn now to the long-time behavior.  Asymptotically, the main
contribution to the second integral in the right-hand side of Eq.~(16)
is gained near the upper limit,
and this integral is easily estimated:
\begin{equation}
\int^t d\tau
e^{v\tau^2/2}\phi_+(\tau) \simeq (vt)^{-1}e^{vt^2/2}\phi_+(t).
\end{equation}
Substituting this estimate into Eq.~(16) and integrating over the
velocity, we arrive at the rate equation,
$d\phi_+/dt=-p_+\phi_+/t$. As a result, the leading asymptotic behavior
for arbitrary input distribution $P_0(v)$ is
\begin{equation}
\phi_{\pm}(t)\sim t^{-p_{\pm}},\qquad t\to \infty.
\end{equation}
This behavior is remarkable, the one-sided problem exhibits
non-universal decay kinetics which is characterized by a simple
parameter, the fraction of right (left) moving seeds nucleating per
unit time.  While the decay for the uncovered fraction for both of the
one-sided problems depends on the initial conditions, and specifically
on the fraction of left and right growing seeds, the uncovered
fraction exhibits a robust decay
\begin{equation}
\Phi(t)\sim t^{-1},\qquad t\to \infty.
\end{equation}
The asymptotic uncovered fraction, $\Phi$, is independent of the input
distribution, and is reminiscent of the temporal behavior in random
sequential adsorption processes \cite{Evans}.  The situation is in
contrast with the heterogeneous case where the presence of slow
particles reduces the asymptotic covering rate. In the next section, we
will show that this robust behavior also emerges from a simple
mean-field theory.

The above analysis enables calculation of additional kinetic
properties of the system such as $n_{s}(t)$, the seed density.
The seeds creation rate is equal to the available space,
and therefore,
\begin{equation}
{dn_s\over dt}=\Phi.
\end{equation}
Using the asymptotic behavior of $\Phi\sim t^{-1}$, we learn that the
seed density grows logarithmically in time,
\begin{equation}
n_s\sim \ln t,   \qquad t\to \infty.
\end{equation}

The velocity distribution function can be calculated following the
same line of reasoning that led to Eq.~(16). Denoting by $P(v,t)$ the
density of growing rays at time $t$, one finds that for $v>0$
\begin{equation}
P_{\pm}(v,t)=P_{\pm}(v)\int_0^t d\tau
\exp[-|v|(t^2-\tau^2)/2]\phi_{\pm}(\tau)\phi_{\mp}(t)
\end{equation}
The integration is carried over all possible creation time $\tau$,
$0<\tau<t$.  As in Eq.~(16), the exponential term ensures that (i) the
interval covered by the growing ray remains empty during the time
interval $(0,\tau)$, and (ii) the space covered by the ray remains
empty during the time interval $(\tau,t)$.  The factor
$\phi_+(\tau)\phi_-(t)$ ensures that the initial position of the seed
and the final position of the tip belong to uncovered area.

In the long-time limit, the main contribution to the integral
in the right-hand side of Eq.~(23) is accumulated near the upper limit.
Thus we can use the estimate of Eq.~(18) to find
\begin{equation}
P(v,t)\simeq {P_0(v)\over|v|}\,{\Phi(t)\over t}.
\end{equation}
The ray velocity distribution is therefore proportional to
$|v|^{-1}P_0(v)$ in the late stages of the process, i.e. slow
velocities are slightly more favorable. However, the relative
enhancement of slow rays is weak in comparison with the heterogeneous
case where fast velocities are exponentially suppressed.  For all
input distributions with finite $\langle |v|^{-1}\rangle_0$ moment, a
universal decay of the density of growing rays is found,
$n_r\simeq\langle |v|^{-1}\rangle_0 \Phi t^{-1}$.  Combining this
result with Eq.~(20), we see that the ray density exhibits universal
decay
\begin{equation}
n_r\sim t^{-2}, \qquad {\rm when} \quad t\to\infty,
\end{equation}
for input distributions with finite $\langle |v|^{-1}\rangle_0$.

Similar to the heterogeneous case, it is useful to consider the power
law distribution, $P_0(v)\sim |v|^{\mu}$ for $|v|\to 0$, with
$\mu>-1$. For this distribution, the moment $\langle
|v|^{-1}\rangle_0$ does not exist when $\mu\leq 0$, and we cannot use
Eq.~(24) in deriving the density of growing rays. Thus we substitute
the exact expression of Eq.~(23) into the relation $n_r=\int dv P(v,t)$,
{\it first} perform $v$-integration, and then $\tau$-integration. This
yields

\begin{equation}
n_r\sim \cases{t^{-2\mu-2}, \quad &$-1<\mu<0$,\cr
               t^{-2}\ln t, \quad &$\mu=0$,\cr
               t^{-2},      \quad &$\mu>0$.\cr}
\end{equation}
The typical ray velocity, defined via
$\bar v=\int dv |v|P(v,t)/\int dv P(v,t)
\equiv \langle|v|\rangle/ n_r$, has the following limiting behavior:
\begin{equation}
\bar v\sim \cases{t^{2\mu},     \quad  &$-1<\mu\leq0$,\cr
                  1/\ln t, \quad  &$\mu=0$,\cr
                  {\rm const},  \quad  &$\mu>0$.\cr}
\end{equation}
Thus the average velocity decreases when the moment $\langle
|v|^{-1}\rangle_0$ is infinite while for all other cases the typical
velocity reaches a limiting value.  To summarize, despite the
universal asymptotic behavior found for the coverage and the seed
density, the ray density and the velocity distribution exhibit
nonuniversal behavior. For the ray density the general behavior is
$n_s\sim t^{-2}$, and only ``pathological'' distributions with enough
slow rays lead to slower ray density decays.

Another interesting quantity is $\rho(l,t)$, the distribution of
frozen rays of length $l$ at time $t$. This distribution can be
readily derived from $P(v,\tau_1,\tau_2)$, the density at time
$\tau_2$ of $v$ rays that were nucleated at time $\tau_1$,
\begin{equation}
\rho(l,t)=-\int dv \int_0^t d\tau_1\int_0^{\tau_1} d\tau_2
{\partial P(v,\tau_1,\tau_2)\over \partial \tau_2}
\delta\left(l-|v|(\tau_1-\tau_2)\right).
\end{equation}
Here, the loss rate of growing rays at time $\tau_1$ is equal to the
gain rate of frozen rays, and the delta function ensures proper length
of the ray.  The conditional density $P(v,\tau_1,\tau_2)$ is given by
the integrand of Eq.~(23), with the transformation $t\to\tau_1$ and
$\tau\to\tau_2$,
\begin{equation}
P_{\pm}(v,\tau_1,\tau_2)=P_{\pm}(v)\,e^{-|v|(\tau_1^2-\tau_2^2)/2}\,
\phi_{\pm}(\tau_1)\phi_{\mp}(\tau_2).
\end{equation}
The jamming distribution can be found by setting $t=\infty$. Combining
the above two equations gives
\begin{eqnarray}
\rho_{\infty}(l)&=&\int dv P_+(v)e^{-l^2/2v}\int_0^\infty d\tau_2
e^{-l\tau_2} \phi_-(\tau_2)\left[\tau_1\phi_+(\tau_1)
+{d\phi_+(\tau_1)\over d\tau_1}\right]_{\tau_1=\tau_2+l/v}\nonumber\\
&+&\int dv P_-(v)\cdots
\end{eqnarray}
The second term is written by exchanging $+$ and $-$. We have not been
able to compute the general behavior. However, we obtain below results in
the special case of bimodal-velocity distributions.

We present now explicit expressions for the both kinetic and the
jamming properties for the bimodal velocity distribution,
$P_0(v)=p_+\delta(v-1)+p_-\delta(v+1)$. Differentiating Eq.~(16) with
respect to $t$ produces the following ordinary differential equation
for $\phi_{\pm}$,
\begin{equation}
{d^2\phi_{\pm}\over dt^2}+t{d\phi_{\pm}\over dt}+p_{\pm}\phi_{\pm}=0.
\end{equation}
This equation is solved subject to the initial conditions
$\phi_{\pm}|_{t=0}=1$ and $d\phi_{\pm}/dt\big|_{t=0}=0$. The solution is
expressed through parabolic cylinder functions [6]
\begin{equation}
\phi_{\pm}(t)=a_{\pm}\exp\left(-t^2/4\right)\left[D_{-p_{\mp}}(t)+
D_{-p_{\mp}}(-t)\right],
\end{equation}
with $a_{\pm}=2^{-1+p_{\mp}/2}\pi^{-1/2}\Gamma(1/2+p_{\mp}/2)$.
The asymptotic behavior agrees with the above analysis,
\begin{equation}
\phi_{\pm}(t)\sim c_{\pm}t^{-p_{\pm}},
\end{equation}
with $c_{\pm}=2^{-p_{\pm}/2}\Gamma(1/2+p_{\mp}/2)/\Gamma(p_{\mp})$.
The uncovered fraction is evaluated from
$\Phi=\phi_+\phi_-$. Thus, the leading asymptotic behavior for
the uncovered fraction is $\Phi(t)\sim c_+c_-t^{-1}$. The asymptotic
velocity distribution is found from Eq.~(24),
\begin{equation}
P(v,t)=c_+c_-\Big(p_+\delta(v-1)+p_-\delta(v+1)\Big)t^{-2}.
\end{equation}
Moreover, for this special case the time dependent velocity
distribution is proportional to the input velocity distribution at all
times. Also, the seed density and the ray density are given by
$n_s\sim c_+c_-\ln t$, and $n_r\sim c_+c_- t^{-2}$, in agreement with
the above theory. Note that the relation $dn_r/dt=-d\Phi/dt$ is
satisfied since the ray speed distribution is monodisperse.

For such a simple velocity distribution it is possible to obtain
several properties of the length distribution in the jammed
configuration. Evaluation of Eq.~(30) using the corresponding
limiting behaviors
$\phi_{\pm}(0)=1$, and $\phi_{\pm}\sim c_{\pm}t^{-p_\pm}$ when $t\to
\infty$, gives
\begin{equation}
\rho_\infty(l)\simeq \cases{c_+c_-\,l^{-1}\left(\ln(1/l)-\gamma\right)
\quad &$l\ll 1$, \cr
(p_+c_+l^{-p_+}+p_-c_-l^{-p_-})\,\exp\left(-l^2/2\right)
\quad &$l\gg 1$. \cr}
\end{equation}
In the above equation $\gamma\cong0.5772$ is the Euler constant.
In the small-size limit, the jammed distribution $\rho_\infty(l)$
exhibits very weak nonintegrable singularity. One can compute the
density of the total number of frozen rays, $F(\epsilon)$, of lengths
greater than $\epsilon$:
\begin{equation}
F(\epsilon)=\int_\epsilon^\infty dl \rho_\infty(l)
\simeq c_+c_-\,\ln^2(1/\epsilon).
\end{equation}
A power-law behavior of the form $F(\epsilon)\sim \epsilon^{-D_f}$
would indicate that $D_f$ is the fractal dimension of the pore space
of forming pattern. Thus, in the present case $D_f=0$, although a weak
logarithmic singularity still appears.

\section{Mean-Field Approximation and Higher Dimensions}

It is worthwhile to consider possible generalizations of the covering
process in higher dimensions.  One such generalization [5] assumes
that the growing objects are rigid spheres whose radius grows
ballistically until a collision occurs. The heterogeneous case is
characterized by a superexponential decay of the covered space
$\Phi(t)-\Phi_{\infty}\sim \exp(-vt^d)$. Another natural
generalization is to objects which do not cover any volume. For
example a growing line in 2D, a growing plane in 3D, {\it etc.}  In
this section we study the growth kinetics of such covering processes
using approximate (mean-field) equations. For one-dimensional
heterogeneous nucleation, the kinetics is highly sensitive to the
details of the velocity distribution, suggesting that mean-field type
theories fail to describe the kinetics. Hence, we focus on the
homogeneous case, where at least in 1D robust behavior was found.

Although the exact solution was obtained in the previous section, it
is instructive to study the covering process using an approximate
approach. We assume a monodisperse velocity distribution
$P_0(v)=\delta(v-1)$, and thus the covering rate is proportional to
the density $n_r$ of growing rays, $d\Phi/dt=-n_r$.  The ray density
itself is estimated from $dn_r/dt=\Phi-n_r(1-\Phi)/l$. Here, rays are
gained with a rate equal to the fraction of uncovered space. The
estimated loss rate is proportional to the density $1-\Phi$ and
inversely proportional to the average length of a gap $l\sim
\Phi$. Hence, the uncovered fraction can be estimated from the
differential equation
\begin{equation}
{d^2\Phi\over dt^2}+(\Phi^{-1}-1){d\Phi\over dt}+\Phi=0
\end{equation}
with the usual initial conditions $\Phi|_{t=0}=1$, and
$d\Phi/dt|_{t=0}=0$.  In fact, both limiting behaviors predicted by
this approximation agree with the exact solution. In the early stages
of the covering process, $\Phi(t)\cong 1-t^2/2$, in perfect agreement
with (17).  When $t\to\infty$, the second derivative term is
negligible and $\Phi$ can be estimated from $d\Phi/dt=-\Phi^2$.
Indeed, the familiar $t^{-1}$ is found for the uncovered
fraction. Also, the total density of growing rays can be found from
$n_r=-d\Phi/dt$, and the resulting $n_r\sim t^{-2}$ asymptotic
behavior is in agreement with Eq.~(25). Despite the success of the
mean-field approximation, the homogeneous covering process was
characterized by nontrivial behavior of the auxiliary one-sided
problem, behavior which can not be accounted for by such a simplified
theory.

Let us now consider the proposed generalization to higher
dimensions. In 2D, The process is defined as follows: Seeds are
nucleated with unit rate in free space. A line grows with unit
velocity from each seed in a random directions until it collides with
another line.  Similarly, in arbitrary dimension $d$, a $d-1$
dimensional hyperplane grows in a random direction and
stops upon collisions.  (By direction, one calls a normal to the
hyperplane). For $d>1$, zero volume is covered and thus, the
seed density is given by $n_s=t$. As the process continues,
hyper-polygons are created. We assume that the density of such
polygons is proportional to the seed density, $n_{p}\sim n_s\sim t$.
Hence, the typical linear size of such objects is $l\sim
n_p^{-1/d}\sim t^{-1/d}$. Finally, the growing objects density
satisfies the generalization of Eq.~(37).
\begin{equation}
\dot n=1-{n n_s\over l}.
\end{equation}
The gain term equals unity since no volume is covered by the
growth. Solving Eq.~(38) we find that asymptotically
\begin{equation}
n\sim t^{-(d+1)/d},\qquad t\to \infty.
\end{equation}

In the limit of infinite dimension, a simple $t^{-1}$ behavior is
found.  In two dimension, the process is equivalent to an isotropic
fragmentation process. A numerical simulation in two dimension with
infinite growth velocity found evidence that $l\sim t^{-1/2}$
\cite{Krapivsky}, in agreement with the above approximation.

\section{Discussion}

We have studied a space covering process in one dimension.  Exact
results for the kinetics and the structure of the system have been
presented for both homogeneous and heterogeneous realizations of the
process. While for the heterogeneous case the temporal behavior of the
coverage is very sensitive to the presence of slowly growing rays, the
asymptotic coverage in the homogeneous case is almost independent of
the details of the input. In both cases the fraction of left and right
moving rays plays a crucial role. In the heterogeneous case, the
jamming coverage depends on these fractions, while conditional coverage
probabilities in the homogeneous case exhibit a surprising algebraic
dependence on these fractions. We also treated a generalization to
higher dimension using of a mean-field theory.

It would be interesting to establish a relationship between the present
process and the chemical processes with ballistically moving
aggregating/annihilating particles.  A natural question is
 whether the nonuniversal behavior found for the one-sided
covering process occurs for systems of interacting ballistically
moving particles.

\section*{Acknowledgment}

E.~B.~ was supported in part by NSF under Award Number 92-08527 and by the
MRSEC Program of the National Science Foundation under Award Number
DMR-9400379.


\begin{thebibliography}{99}

\bibitem{Bray}
     A.~J.~Bray, {\sl Adv. Phys.}   {\bf 43}, 357 (1994).
\bibitem{Sekimoto}
     K.~Sekimoto,  {\sl Int. J. Mod. Phys. B} {\bf 5}, 1843 (1991).
\bibitem{Evans}
     J.~W.~Evans,  {\sl Rev. Mod. Phys.} {\bf 65}, 1281 (1993).
\bibitem{Meakin}
     P.~Meakin, {\sl Rep. Prog. Phys.} {\bf 55}, 157 (1992).
\bibitem{Privman}
     M.~C.~Bartelt and V.~Privman,  {\sl Int. J. Mod. Phys. B} {\bf 5},
     2883 (1991).
\bibitem{Viscek}
     T.~Viscek, {\it Fractal Growth Phenomena}
     (World Scientific, Singapore, 1992).
\bibitem{sa}
     D.~Stauffer and A.~Aharony, {\it Introduction to Percolation Theory}
     (Taylor and Francis, London, 1992).
\bibitem{Jager}
     W.~Jager, H.~Rost, and P.~Tautu, {\it Biological Growth and Spread}
     Lecture Notes in Mathematics, Vol.~38 (Springer, Berlin).
\bibitem{Bradley}
     R.~M.~Bradley and P.~N.~Strenski, {\sl Phys. Rev. B}  {\bf 40},
     8967 (1989).
\bibitem{Andrienko}
     Yu.~A.~Andrienko, N.~V.~Brilliantov and P.~L.~Krapivsky,
     {\sl Phys. Rev. A} {\bf 45}, 2263 (1992); P.~L.~Krapivsky, {\sl
     J. Chem. Phys.} {\bf 97}, 8817 (1992).
\bibitem{Brilliantov}
     N.~V.~Brilliantov, P.~L.~Krapivsky, and Yu.~A.~Andrienko,
     {J. Phys. A} {\bf 27}, L381 (1994).
\bibitem{Krapivsky}
     P.~L.~Krapivsky and E.~Ben-Naim, {\sl Phys. Rev. E} {\bf 50},
     3502 (1994).
\bibitem{india}
     M.~Rao, S.~Dasgupta, and H.~K.~Sahu, {\sl Phys. Rev. Lett.} {\bf 75},
     2164 (1995).
\bibitem{Bender}
     C.~M.~Bender and S.~A.~Orszag, {\it Advanced Mathematical
     Methods for Scientists and Engineers} (McGraw-Hill, Singapore, 1984).
\bibitem{Ben-Naim}
     E.~Ben-Naim, S.~Redner, and F.~Leyvraz, {\sl
     Phys. Rev. Lett.} {\bf 70}, 1890 (1993);
     E.~Ben-Naim, P.~L.~Krapivsky,
     and S.~Redner, {\sl Phys. Rev. E} {\bf 50}, 822 (1994).

\end{thebibliography}
\end{document}